\documentclass[a4paper,11pt]{article}
\pdfoutput=1 

\usepackage{jheppub} 
                     
\usepackage{amsmath,amssymb,amsthm,amscd,graphicx}
\usepackage{psfrag}
\usepackage[english]{babel}
\usepackage{float}
\input epsf.sty

\usepackage{color}

\theoremstyle{definition}

\newcommand{\CE}{{\cal E}}

\newcommand{\CN}{{\cal N}}
\newcommand{\CO}{{\cal O}}

\newcommand{\CV}{{\cal V}}
\newcommand{\CW}{{\cal W}}

\newcommand{\mO}{{\mathsf{O}}}

\def\IN{{\mathbb N}}
\def\IZ{{\mathbb Z}}
\def\IR{{\mathbb R}}
\def\IQ{{\mathbb Q}}
\def\IC{{\mathbb C}}
\def\IP{{\mathbb P}}

\def\IS{{\mathbb S}}
\def\IF{{\mathbb F}}

\newcommand{\re}{{\rm e}}
\newcommand{\ri}{{\rm i}}
\newcommand{\rd}{{\rm d}}

\newcommand{\mx}{{\mathsf{x}}}

\newcommand{\mm}{\mathsf{p}}

\newcommand{\be}{\begin{equation}}
\newcommand{\ee}{\end{equation}}
\newcommand{\ba}{\begin{aligned}}
\newcommand{\ea}{\end{aligned}}
\newcommand{\ben}{\begin{eqnarray}\displaystyle}
\newcommand{\een}{\end{eqnarray}}

\newdimen\tableauside\tableauside=1.0ex
\newdimen\tableaurule\tableaurule=0.4pt
\newdimen\tableaustep
\def\phantomhrule#1{\hbox{\vbox to0pt{\hrule height\tableaurule width#1\vss}}}
\def\phantomvrule#1{\vbox{\hbox to0pt{\vrule width\tableaurule height#1\hss}}}
\def\sqr{\vbox{%
  \phantomhrule\tableaustep
  \hbox{\phantomvrule\tableaustep\kern\tableaustep\phantomvrule\tableaustep}%
  \hbox{\vbox{\phantomhrule\tableauside}\kern-\tableaurule}}}
\def\squares#1{\hbox{\count0=#1\noindent\loop\sqr
  \advance\count0 by-1 \ifnum\count0>0\repeat}}
\def\tableau#1{\vcenter{\offinterlineskip
  \tableaustep=\tableauside\advance\tableaustep by-\tableaurule
  \kern\normallineskip\hbox
    {\kern\normallineskip\vbox
      {\gettableau#1 0 }%
     \kern\normallineskip\kern\tableaurule}%
  \kern\normallineskip\kern\tableaurule}}
\def\gettableau#1{\ifnum#1=0\let\next=\null\else
\squares{#1}\let\next=\gettableau\fi\next}

\newcommand*{\tabref}[1]{\tablename~\ref{#1}}

\newcommand{\figref}[1]{Fig.~\protect\ref{#1}}

\title{\boldmath The complex side of the TS/ST correspondence}

\author{ Alba Grassi$^a$ and Marcos Mari\~no$^b$}

\affiliation{
$^a$International Center for Theoretical Physics,\\
 ICTP, Strada Costiera 11, Trieste 34151, Italy \\
 \\
 $^b$ D\'epartement de Physique Th\'eorique et Section de Math\'ematiques,\\
Universit\'e de Gen\`eve, Gen\`eve, CH-1211 Switzerland}

\emailAdd{agrassi@ictp.it, marcos.marino@unige.ch}

\abstract{The TS/ST correspondence relates the spectral theory of certain quantum mechanical operators, to topological strings on toric Calabi--Yau threefolds. So far the correspondence has 
been formulated for real values of Planck's constant. In this paper we start to explore the validity of the correspondence when $\hbar$ takes complex values. We give evidence that, for threefolds 
associated to supersymmetric gauge theories, one can extend the correspondence and obtain exact quantization conditions for the operators. We also explore the correspondence for operators involving 
periodic potentials. In particular, we study a deformed version of the Mathieu equation, and we solve for its band structure in terms of the quantum mirror map of the underlying threefold.}

\begin{document}
\maketitle

\flushbottom

\section{Introduction}

The topological string/spectral theory (TS/ST) correspondence is a conjectural relationship between topological string theory 
on toric Calabi--Yau (CY) manifolds and the spectral theory of certain trace 
class operators on the real line. This correspondence was built on previous work in \cite{adkmv,ns,mirmor,acdkv,mp,hmo,hmo2,calvom,hmo3,hmmo,km,hw,cgm8}
formulated in detail in \cite{ghm,cgm} (see \cite{mmrev} for a review), and further developed in 
\cite{kama,mz,hatsuda, kmz,wzh,gkmr, hatsuda-comments, lst, hm, fhm, oz, bgt, ag, kpamir, Sciarappa-1,sug,butterfly,mz-wv, swh, cgum,ggu,cms,ghkk,hsx,bgt2,Sciarappa-2,mz-wv2}. 
The operator appearing in the correspondence is obtained by quantization of the mirror curve to the toric CY, as originally envisaged in \cite{adkmv}. The TS/ST correspondence 
provides, among other things, explicit and exact quantization conditions for the spectrum of the operator, in terms of BPS invariants of the CY manifold, as well as a non-perturbative 
definition of the topological string partition function. 

So far, studies of the TS/ST correspondence have focused on the ``physical" case in which $\hbar$ is real. In addition, the quantization of the mirror curve is done along a real slice. As long as appropriate positivity constraints are imposed on some of the parameters, the resulting operator is self-adjoint and trace class, and it has a discrete, real spectrum. 
From the point of view of operator theory, this is the simplest situation. However, it is natural to explore other possibilities, in which the quantities specifying the model are allowed 
to take more general values, or the quantization of the curve is done with a different prescription. For example, in \cite{cgum} the positivity constraints on some of the parameters 
were relaxed, and this led to resonant-like 
states in the spectrum. 

In this paper we consider two additional cases. We first study the case in which 
$\hbar$ is allowed to take complex values. When this happens, the operators obtained by quantization of the mirror curve are no longer 
self-adjoint, but we can still make sense of their spectral problem (in particular, their spectrum can be computed numerically). In this paper we give 
evidence that the exact quantization condition of \cite{ghm,wzh} still captures the exact spectrum of the operator when the underlying
 toric CY engineers a five-dimensional gauge theory. This requires a partial resummation of the BPS expansion which is natural from the gauge theory point of view (such a resummation was first considered in topological string theory in \cite{ikp, ikp3} to reproduce Nekrasov's results \cite{n} from the 
topological vertex \cite{akmv}). 

In addition, we consider a mixed quantization of the mirror curve, in which one of the coordinates becomes purely imaginary. 
For simplicity, we focus on the operator associated to local $\IF_0$, 
which provides a quantum, or deformed version, of the Mathieu equation. Since this operator involves a periodic potential, there is a band structure for the eigenvalue problem 
which can be analyzed with standard tools. Our main result is an exact expression for the band energies in terms of a resummation of the 
quantum mirror map. This generalizes well-known results for the Mathieu equation \cite{ns,he-miao,basar-dunne,du-mathieu}. 

This paper is organized as follows. In section 2 we review the basics of the TS/ST correspondence. In section 3 we consider the correspondence for complex values of Planck's constant, and 
we study in detail the spectral problem associated to local $\IF_0$. In section 4 we analyze the band spectrum of the quantum Mathieu operator. Finally, in section 5 we conclude and list some 
open problems. The Appendix gives some technical details on the grand potential of a local CY threefold. 

\section{Topological strings and spectral problems } \label{s2}

In this paper we are interested in spectral problems arising in the quantization of mirror curves to toric CY manifolds.  Let us start by recalling some facts concerning this quantization.
We consider a  toric CY $X$ whose complexifed K\"ahler moduli space is parametrised by $g_\Sigma$ ``true" moduli 
\be\label{tm} \kappa_i, \quad i=1, \cdots, g_\Sigma
 \ee
and $r_\Sigma$ mass parameters
\be
\label{mass} \xi_i, \quad i=1, \cdots, r_\Sigma. \ee
We will denote by $n_\Sigma=g_\Sigma+r_\Sigma$ the total number of K\"ahler parameters. 
A more precise definition of these two types of moduli can be found in \cite{hkp,hkrs}.
The mirror curve to $X$ has genus $g_\Sigma$, and there are $g_\Sigma$ different ``canonical" forms to represent it, namely
\be \mathcal{O}_i (x,p)+\kappa_i=0, \quad i=1, \cdots, g_{\Sigma},\ee
where $ \mathcal{O}_i (x,p)$ is a sum of monomials of the form $\re^{ax+bp}$. The coefficients of such monomials depend on the mass parameters \eqref{mass} and on the moduli 
$\kappa_j$, $j\neq i$. 
We promote the variables $x,p$ to operators $ \mathsf x,\mathsf p $ such that
\be \label{quant} [\mathsf x, \mathsf p]=\ri \hbar. \ee
By using Weyl quantization
\be \re^{ax+bp}\to \re^{a \mathsf x+b \mathsf p} \ee
we can promote $\mathcal{O}_i (x,p)$ to an operator ${\mathsf O}_i $, $i=1, \cdots, g_\Sigma$. 
For example, the mirror curve to the canonical bundle over $\IF_0=\IP^1 \times \IP^1$ has genus one and one mass parameter $\xi$, and we have
\be \mathcal{O}_1(x,p)+\kappa= \re^{x}+\xi \re^{-x}+\re^{p}+ \re^{-p} +\kappa.\ee
The corresponding operator is
\be \label{opp1p1} {\mathsf O}_1= \re^{\mathsf x}+\xi \re^{- \mathsf x}+\re^{\mathsf p}+ \re^{- \mathsf p} .\ee
 It was conjectured in \cite{ghm}, and proved in \cite{kama,lst} for many examples, that the inverse operators
 \be \label{rho}  { \rho}_i={\mathsf O}_i^{-1}\ee
 are  self-adjoint, positive and of trace class, provided some positivity constraints are imposed on the mass parameters. Therefore, 
 they have a discrete spectrum which is encoded in the generalised spectral determinant \cite{cgm}
 \be \label{sd}\Xi_X({\boldsymbol \kappa}, \hbar, {\boldsymbol \xi})=\det \left(1+\sum_{j=1}^{g_\Sigma} {\mathsf A}_{ij} \kappa_j\right)  \ee
 where the operators ${\mathsf A}_{ij}$ are defined by
 \be{\mathsf O}_i+\kappa_i={\mathsf O}_i^{(0)}\left(1+\sum_{j=1}^{g_\Sigma} {\mathsf A}_{ij} \kappa_j\right), 
 \ee
 and  ${\mathsf O}_i^{(0)}$ does not depend on the true moduli $\kappa$ (see \cite{cgm} for more details).
 The conjecture of \cite{ghm,cgm} states that
 \be \label{conj} \Xi_X({\boldsymbol \kappa},\hbar,{\boldsymbol \xi})=\sum_{{\bf n} \in \IZ^{g_\Sigma} }\re^{\mathsf{J}_X({\boldsymbol \mu}+2 \pi \ri {\bf n},\hbar,{\boldsymbol \xi})}, \quad \kappa_i=\re^{\mu_i}, \quad \hbar \in \IR_+\ee
 where $\mathsf J_X$ is the topological string grand potential \cite{hmmo} and it is completely determined by the (refined) BPS invariants of $X$. The precise definition of 
 $\mathsf J_X$ is given in appendix \ref{defJ}. This construction, and in particular \eqref{conj}, has led to a new and exact relation between topological strings 
 and the spectral theory of the quantum operators arising in the quantization of mirror curves. We will refer to it as TS/ST correspondence, and it has 
 passed a large number of successful tests.
 
 In this paper we  focus only  on one particular consequence of this correspondence which concerns the quantisation condition 
 for the spectrum of the operators $\rho_i$.  As explained in \cite{ghm,cgm} the eigenvalues of these operators are determined by the vanishing locus of the 
 spectral determinant \eqref{conj}. In particular the WKB part of the spectrum is encoded in the NS limit of the refined topological string, as already anticipated in \cite{ns,acdkv,mirmor, Bonelli:2011na}. However there are additional non-perturbative corrections which are encoded in the unrefined (or GV) limit of topological string theory \cite{km,ghm,cgm}. 
 When the mirror curve has genus one, by using the blowup equations \cite{naga}, it is possible to express the vanishing locus of $\Xi_X({ \kappa},\hbar,{\boldsymbol \xi})$ in a way which displays S-duality \cite{wzh,swh,ggu}. More precisely, the blowup equations allow to write the vanishing condition
 \be \label{xi0}\Xi_X( \kappa,\hbar,{\boldsymbol \xi})=0\ee
as
 \be \label{g1ns}   \sum_{i=1}^{n_\Sigma} c_i \left( \frac{\partial}{\partial t_i} \widehat{F}^{\rm NS}(\widehat{\mathbf{t}}(\hbar), \hbar) + \frac{\hbar}{2\pi}\frac{\partial}{\partial t_i} \widehat{F}^{\rm NS,inst}\left(\frac{2\pi}{\hbar} \widehat{\mathbf{t}}(\hbar), { 4 \pi^2 \over  \hbar}\right) \right) =n+{1\over 2}, \quad  n= 0,1, \cdots  \ee
In this equation, the coefficients $c_i$ are given by $c_i=C_{i1}$, where the matrix $C_{ij}$ is defined in \eqref{zmu}, $\widehat{F}_{\rm NS}, \widehat{F}_{\rm NS, inst}$ denote the twisted NS free energy \eqref{NS-j-hat}, \eqref{NS-inst-hat}, and $\widehat{t}_i(\hbar)$ is the twisted quantum mirror map \eqref{qmirhat}.  Equation \eqref{xi0}, or equivalently \eqref{g1ns}, should be viewed  as a quantization condition for the complex modulus $\kappa$, and computing the spectrum of $\rho_1$. 
Let us denote by 
\be \{\re^{-E_n}\}_{n\geq 0}
\ee
 the eigenvalues of $\rho_1$. Then, 
 \be 
 \label{sp}  \{-\re^{E_n}\}_{n\geq 0}= \{ \kappa^{(n)} : \quad \Xi_X( \kappa^{(n)},\hbar,{\boldsymbol \xi}) =0\}.\ee  

 The higher genus situation is more subtle and there are in principle two different spectral problems associated to a given CY.
 The first one was studied in \cite{cgm}, where one considers the $ g_\Sigma$ non-commuting operators $\mO_i$, $i=1, \cdots, g_\Sigma$ 
 acting on $L^2(\IR)$. The spectrum of these operators is encoded in the vanishing locus of the generalized spectral determinant \eqref{sd}, 
 which defines a codimension one submanifold in the moduli space.
 The second one is the spectral problem associated to the quantum cluster integrable system of Goncharov and Kenyon \cite{gk}. In this case, one has $g_\Sigma$ 
 commuting hamiltonians acting on  $L^2(\IR^{g_\Sigma})$, whose spectrum is determined by the following $g_\Sigma $ exact quantization conditions \cite{hm,fhm}
  \be 
  \label{gns}  
  \sum_{i=1}^{n_\Sigma} C_{ji}\left( \frac{\partial}{\partial t_i} \widehat{F}^{\rm NS}(\widehat{\mathbf{t}}(\hbar), \hbar) + \frac{\hbar}{2\pi}\frac{\partial}{\partial t_i} \widehat{F}^{\rm NS,inst}\left(\frac{2\pi}{\hbar} \widehat{\mathbf{t}}(\hbar), { 4 \pi^2 \over  \hbar}\right) \right) =n_j+{1\over 2},     \quad j=1, \cdots, g_\Sigma,
   \ee
 where $n_j= 0,1, \cdots$ are non-negative integers. It was pointed out in \cite{hm,fhm}, and verified in \cite{swh,ggu}, that the spectral 
 problem of \cite{cgm} is more general than the one of \cite{gk}, in the sense that 
   \eqref{gns} defines a subset of $g_\Sigma$ points lying on the vanishing locus of \eqref{sd}. This point of view was confirmed recently by \cite{mz-wv2} in the study of eigenfunctions.

 \section{Complexifying Planck's constant} 
 \label{s3}

 In the current formulation of the TS/ST correspondence it is always assumed that 
 \be \label{real} \hbar \in \IR_+.\ee 
 One of the reasons for this restriction is that, as first pointed out in \cite{hmmo}, some of the expansions appearing in the construction of \cite{ghm,cgm} 
 {\it diverge} when $\hbar$ is complex. In this section we will see that, by using insights from gauge theory, we can overcome this difficulty and easily extend some 
 aspects of the TS/ST correspondence to complex values of $\hbar$, at least when the underling CY can be use to engineer gauge theories \cite{kkv}. We will focus, for concreteness 
 and simplicity, on the 
 spectral problem associated to local $\IF_0$, which corresponds to the pure $\CN=2$, $U(2)$ gauge theory in 5d. 
 
 \subsection{Convergent expansions } 

Let us first consider the topological string partition function, as computed for example from the topological vertex \cite{akmv}.
This quantity has the following structure:
\be \label{FGvt} Z^{\rm GV}({\bf t},g_s)=\exp[F^{\rm GV}({\bf t},g_s)]=\sum_{{\bf m}} d_{{\bf m}}(q) \re^{-\bf m \cdot t }, \quad q=\re^{\ri g_s} ,
\ee
where $F^{\rm GV}({\bf t},g_s)$ is defined in \eqref{GVgf}, and ${\bf m}$ is a $n_\Sigma$-uple of non-negative integers (by convention, $d_{(0,\cdots, 0)}=1$). 
The coefficients $d_{{\bf m}}$ have poles at $g_s \in \pi \IQ$, and as consequence the partition function is ill-defined on the real axis. When $g_s\in \IC \backslash \IR$ the coefficients  $d_{{\bf m}}$ do not have poles, however the formal power series \eqref{FGvt} diverges \cite{hmmo}.
A similar analysis can be done for the NS partition function \eqref{NS-inst} as shown in \cite{kpamir}. 
 Nevertheless  when the CY $X$ can be used to engineer a 5d gauge theory, it is possible to partially resum \eqref{FGvt}, 
 as done in \cite{ikp3,ta,ikv} by using the insights coming from gauge theory \cite{n}.  
 
 Let us consider for simplicity 5d $U(N)$ gauge theory on $\IS^1\times \IR^4$, 
and let $a_I$ be the parameters for the Coulomb branch, $I=1, \cdots, N$. Let $Y_I$ be the Young tableau describing the contribution of the instanton in the $I$-th factor 
of the gauge group, and let us denote by $|Y_I|$ the total number of boxes in the tableau. We  group the 
different Young tableaux in a vector of partitions
\be
\boldsymbol{Y}=(Y_1, \cdots, Y_N). 
\ee
Then the partition function of the 5d theory can be written as a sum over Young tableaux, 
\be
Z=\sum_{\boldsymbol{Y}} \Lambda^{|\boldsymbol{Y}|} z_{\boldsymbol{Y}}, 
\ee
where $z_{\boldsymbol{Y}}$ is an appropriate function which we will make explicit below for local $\IF_m$ (i.e. for $N=2$), and 
\be
 |\boldsymbol{Y}|= \sum_{I=1}^N |Y_I|. 
 \ee
Let us first list the ingredients needed in order to write down the refined topological string partition function for local $\IF_m$. 
The first ingredient is 
\be
Z_\mu(t, q)= \prod_{(i,j) \in \mu} \left(1- t^{\mu_j^t-i+1} q^{\mu_i-j}\right)^{-1}, 
\ee
where $\mu$ is a partition or Young tableau, and the parameters $q$, $t$ encode the $\epsilon_{1,2}$ deformations:
\be
q=\re^{\ri \epsilon_1}, \qquad t=\re^{-\ri \epsilon_2}. 
\ee
The second ingredient depends on two partitions $\mu$, $\nu$, and an extra parameter $Q$. It is given by 
\be
R_{\mu \nu}(Q)= \prod_{i,j=1}^\infty {1- Q t^{j-1/2} q^{i-1/2} \over 1-  Q t^{j-1/2- \mu_i} q^{i-1/2-\nu_j}}. 
\ee
It is easy to see that the product gets truncated and only a finite number of factors get involved. We also introduced, for a given partition $\mu$, 
the quantities
\be
\ba
|\mu|&=\sum_i \mu_i, \\
\Vert \mu \Vert^2&= \sum_i \mu_i^2,\\
\kappa_\mu&= \sum_i \mu_i(\mu_i-2i+1),
\ea
\ee
and the refined framing factor
\be
f_\mu= (-1)^{|\mu|}\left( {t \over q} \right)^{\Vert \mu^t \Vert^2/2} q^{-\kappa_\mu/2}, 
\ee
where $\mu^t$ denotes the transposed partition in which one exchange rows and columns of the corresponding 
Young diagram. The building block of the partition function is 
\be
Z_{\mu^1, \mu^2}=q^{\sum_{i=1}^2 \Vert \mu^i\Vert^2/2} t^{\sum_{i=1}^2 \Vert \mu^{i,t}\Vert^2/2} \prod_{i=1}^2 Z_{\mu^i} (t,q) Z_{\mu^{i,t}}(q,t)
R_{\mu^{1,t},\mu^2}\left( {\sqrt{t\over q}} Q_2\right)R_{\mu^{1,t},\mu^2}\left( {\sqrt{q\over t}} Q_2\right).
\ee
Then, we have 
\be
z_{(\mu^1, \mu^2)}(Q_2)=\sum_{\mu^i}  f_{\mu^1}^{-m-1} f_{\mu^2}^{-m+1} Z_{\mu^1, \mu^2}  Q_2^{m |\mu^1|}, 
\ee
so that the partition function of the local $\IF_m$ geometry is given by 
\be\label{reftv}
Z_{\IF_m}(Q_1, Q_2,q, t)= \sum_{Y_1, Y_2} z_{(Y_1, Y_2)}\cdot (-Q_1)^{\sum_{i=1}^2 |Y_i|} .
\ee
For instance  for local  $\IF_0$, i.e. $m=0$,  we have
 \be
 \label{Zeg} Z_{\rm \IF_0}(Q_1,Q_2,q, 1)
 =\exp\left[\sum_{n\geq 1}{\frac{q^n+1}{n^2 \left(q^n-1\right)}} Q_2^n\right]\left[1+\sum_{m\geq 1} b_m(Q_2,q) Q_1^m \right], 
 \ee
 where the first coefficient reads
 \be 
 b_1(Q_2,q) =\frac{q \left(q+1\right)}{\left(1-q\right) \left(q-Q_2\right) \left(q Q_2-1\right)}.
 \ee
It was  proven in \cite{bsu} that the formal series in \eqref{reftv} converges in the standard topological string limit
\be
q=t^{-1}=\re^{\ri g_s}
\ee
provided that 
\be \label{conv-conds}
|q| \neq 1 \qquad Q_2 \neq q^n, \quad n \in \IZ. 
\ee
  We expect \eqref{reftv} to converge as well in the NS limit 
 \be
 t=1, \qquad q=\re^{\ri \hbar}
 \ee
  when (\ref{conv-conds}) holds, as one can check with the explicit expansions. 
 Therefore, in order to ensure good convergence properties for complex values of the string coupling constant in the standard topological 
 string limit, or for complex values of the Planck constant in the NS limit, 
 one has to use the resummed expression for the (refined) topological string free energy, as obtained from the 5d gauge theory computation.

Another important ingredient in the TS/ST duality is  the quantum mirror map. This was introduced in \cite{acdkv}
and has the following form
\be \label{qmir2} -t_i(\hbar)=\log z_i+ \Pi_i( \bf z, \hbar), 
\ee
where $z_i$ are the Batyrev coordinates  defined in \eqref{zmu} and $ \Pi_i$ is a power series in $z_j$. When $\hbar$ is real 
this series has a finite radius of convergence as discussed for instance in \cite{hmmo}.
When $\hbar$ is complex, this is no longer the case. Nevertheless, as for the free energy, we can overcome this problem 
by using instanton calculus in toric CY threefolds with a gauge theory realization. 
More precisely, the quantum mirror map can be computed by using Wilson loop vevs in the gauge theory, as pointed out in 
\cite{Sciarappa-1}. We follow \cite{bk,bkk} for the Wilson loop vev computation. 
Let us define
\be
\ba
\CW&= \sum_{I=1}^N \re^{a_I}, \\
\CV_{\boldsymbol{Y}}&=\sum_{I=1}^N \re^{a_I}\sum_{(k,l)\in Y_I}\re^{\ri (l-1) \epsilon_1 +\ri (k-1)\epsilon_2}. 
\ea
\ee
The indices $(k,l)$ label the boxes of the Young tableau. We have changed slightly the formula in \cite{bk,bkk} to fit our own conventions. 
We also need the equivariant Chern character, 
\be
{\rm Ch}_{\boldsymbol{Y}}(\CE)= \CW- (1-\re^{\ri \epsilon_1}) (1-\re^{\ri \epsilon_2}) \CV_{\boldsymbol{Y}}. 
\ee
The vev of a Wilson loop in the fundamental representation is then given by 
\be \label{ws}
W={1\over Z} \sum_{\boldsymbol{Y}} \Lambda^{|\boldsymbol{Y}|} {\rm Ch}_{\boldsymbol{Y}}(\CE) z_{\boldsymbol{Y}}.
\ee
It turns out that this can be used to compute the quantum mirror map, at least in the simple case of $U(2)$ gauge theories, where there is only one non-trivial 
mirror map (for theories of higher rank, one probably has to use Wilson loops in higher representations). 

Let us look at the explicit form of \eqref{ws} for local $ \IP^1\times \IP^1$. In this case, the ingredients for the Wilson loop give, when expressed in terms of the natural exponentiated K\"ahler parameters, 
\be
\ba
\CW&=Q_2^{1/2}+ Q_2^{-1/2},\\
{\rm Ch}_{\boldsymbol{Y}}(\CE)&=Q_2^{1/2}+ Q_2^{-1/2}\\
&- (1-q)(1-t^{-1})\left( Q_2^{1/2}\sum_{(k,l)\in Y_1}q^ {l-1}t^{-k+1}+Q_2^{-1/2}\sum_{(k,l)\in Y_2}q^ {l-1}t^{-k+1}\right).
\ea
\ee
We find, in the NS limit, a result of the form
\be
Q_2^{1/2} W(Q_1,Q_2, q)= 1+ Q_2+ {(1+ Q_2) Q_1 \over \left(q^{1/2} Q_2- q^{-1/2}\right)\left( q^{-1/2} Q_2 - q^{1/2} \right)}+ \cdots
\ee
which can be re-expanded in powers of $Q_2$, $Q_1$:
\be
\ba
Q_2^{1/2} W(Q_1,Q_2, q)&= 1+Q_2+ Q_1 + (q^{-1}+ 1 + q) Q_2 Q_1\\
& + (q^2+ q+1+q^{-1}+ q^{-2}) Q_2 Q_1 (Q_2+ Q_1)+ \cdots
\ea
\ee
We note that 
\be
Q_2= \xi Q_1,  \quad Q_1=\re^{-t_1(\hbar)}, \quad q=\re^{\ri \hbar},
\ee
where $\xi$ is a mass parameter and $t_1(\hbar)$ is the quantum mirror map \eqref{qmir}. We can write 
\be\label{WQ}
  W^2(Q_1,\xi Q_1, q)= Q_1^{-1} \xi^{-1}\left(1+(2 \xi +2) Q_1+Q_1^2 \left(\xi ^2+4 \xi +2 \xi  q+\frac{2 \xi }{q}+1\right)+\cdots\right).
\ee
It follows that $W$ can be identified with 
\be \label{wmu}
 W^2(Q_1,\xi Q_1, q)={ 1\over { \xi z }}, 
\ee
where $z=\re^{-2 \mu}$ is the bare modulus of the CY, which can be expressed in terms of $Q_1$ and $\xi$ by using the quantum mirror map. 
This can be checked against explicit calculations. 

Let us note that $ W(Q_1,Q_2, q)$ has poles when $Q_2=q^{\pm m}$ for $m \in \IN$. However, these poles do not occur at the values of the 
K\"ahler parameters selected by the quantization condition. More precisely, if $t_1(\hbar, \xi)$ satisfies 
\eqref{gns} for some fixed value of  $\hbar, \xi$, then 
\be Q_2= \xi \re^{-t_1(\xi, \hbar)} \neq \re^{ \pm m\ri \hbar}.\ee
As we will see in the next section, in the case of the quantum Mathieu operator, this is no longer the case and these poles occur at the edges of the energy bands.

 \subsection{Exact quantization condition for complex $\hbar$}

Given the observations above, it is natural to expect that the quantization condition of \cite{ghm}, in the form of \cite{wzh}, 
still holds when $\hbar$ takes complex values, 
provided we consider a partial resummed version of
the free energies and of the quantum mirror map. 
We will test this expectation in the case of local $\IP^1 \times \IP^1$ where the quantization condition 
\eqref{g1ns} can be written as 
\be
\label{EQC}
-2 \left\{ {\partial  \over \partial t_j}  F^{\rm NS}\left ({\boldsymbol{t}}(\hbar), \hbar \right) +
\frac{\hbar}{2\pi}
 {\partial \over \partial t_j} F_{\text {NS, inst}}\left({2 \pi \over \hbar} {\boldsymbol{t}}(\hbar), {4 \pi^2  \over \hbar} \right)\right\}= 2 \pi  \left( n+{1\over 2} \right), \quad n=0,1,\cdots.
 \ee
We need to consider the resummed form of $F^{\rm NS}$ given by the NS limit of \eqref{reftv}. The B--field can be set to zero in this geometry \cite{hmmo,ghm}. 
The very first orders of the expansion of (\ref{EQC}) in $Q_1$, $Q_1^D=Q_1^{2 \pi/\hbar}$ are given by
\be\label{EQexp}\ba  &\frac{t_1(\hbar)^2}{\hbar }-\frac{\hbar }{6}-\frac{2 \pi ^2}{3 \hbar }-\frac{t_1(\hbar) \log \xi}{\hbar }+2 { \sum_{w\geq 1}}{1\over w}\left(Q_2^{\frac{2 \pi  w}{\hbar }} \cot \left(\frac{4 \pi ^2 w}{2 \hbar }\right)+Q_2^w \cot \left(\frac{w \hbar }{2}\right)\right)\\
&-\frac{2 Q_1^D \left((Q_2^D)^2-1\right) \cot \left(\frac{2 \pi ^2}{\hbar }\right)}{\left((Q_2^D)^2-2 Q_2^D \cos \left(\frac{4 \pi ^2}{\hbar }\right)+1\right)^2}-\frac{4 \ri Q_1 \left(Q_2^2-1\right) \re^{\frac{5 \ri \hbar }{2}} \cos \left(\frac{\hbar }{2}\right)}{\left(-1+\re^{\ri \hbar }\right) \left(-Q_2+\re^{\ri \hbar }\right)^2 \left(-1+Q_2 \re^{\ri \hbar }\right)^2}+\mathcal{O}(Q_1^2,(Q_1^D)^2 )\\
&= 2 \pi  \left( n+{1\over 2}\right), \quad n=0,1, \cdots\ea \ee
where 
\be Q_i^D=Q_i^{2 \pi/\hbar}, \quad Q_2=\xi Q_1, \quad Q_1=\re^{-t_1(\hbar)}\ee
and $t_1(\hbar)$ is the quantum mirror map \eqref{qmir}.
Therefore, for fixed $\hbar$ and  $\xi$, the quantization condition \eqref{EQC} is an equation for $t_1(\hbar)$. 
Let us denote by \be t_1^{(n)}(\hbar, \xi)\ee the solution to \eqref{EQC} for a fixed value of $\hbar, \xi, n$. 
 It follows that the spectrum   of  \eqref{opp1p1},  denoted by $\re^{E_n}$, is given by
\be \label{WvsE}{ \xi^{-1}}\re^{2 E_n}= \left(W\left(\re^{-t_1^{(n)}(\hbar, \xi)}, \xi \re^{-t_1^{(n)}(\hbar, \xi)}, \re^{\ri \hbar}\right)\right)^2 .\ee
In the following we will assume by simplicity that the mass parameter in (\ref{opp1p1}) is set to one $\xi=1$. 
\begin{table}[H]
\begin{center}
 \begin{tabular}{ccccc}
	
	$n$ & $E_0 $    \\ \hline
	3 &\underline{2.1234991}550043 + \underline{0.238200}0043930 \ri\\ 
	 5&\underline{2.123499196}9099 + \underline{0.238200118}0857 \ri\\ 
	 6&\underline{2.12349919683}06 + \underline{0.23820011838}39 \ri \\ \hline
Num &                    2.1234991968312 +                 0.2382001183808  \ri  &
	     \end{tabular}
        \caption{ The first energy level for the operator \eqref{opp1p1} with $\xi=1$ at $\hbar=3+\ri$ obtained by solving \eqref{EQC} and by using  \eqref{WvsE}. We denote by $n$ the order at which 
        we truncate the $Q_1$-series in \eqref{reftv}. The numerical value is computed  by using matrices of size 300.   }  \label{TaB}
\end{center}
          \end{table}
In  \tabref{TaB} and  \tabref{hTaBb} we list some results for the spectrum of \eqref{opp1p1} with $\xi=1$, which are obtained by solving \eqref{EQC}, for different values of $\hbar$. 
Note that the eigenvalues are now complex, and we order them by their absolute value, i.e. $|E_0|<|E_1|<\cdots$. These eigenvalues can be also 
computed numerically. To do this, we perform a standard diagonalization of the operator in the harmonic oscillator basis, analytically continued 
to complex values of $\hbar$. The numerical results are in perfect agreement with the results obtained from the exact quantization condition 
(\ref{EQC}).

   \begin{table}[H] \begin{center}
         \begin{tabular}{ccccc}
	$n$ & $E_0 $    \\ \hline
	3 &   \underline{2.22473262}2055 ~+ \underline{0.7856673}09664 \ri  \\ 
	 5&   \underline{2.224732626}247 +   \underline{0.7856673858}47 \ri\\ 
	 6&   \underline{2.22473262633}5 +   \underline{0.78566738581}6 \ri \\ \hline
   Num &   2.224732626334 +  0.785667385819      \ri  &
	     \end{tabular}
        \caption{ The first energy level for the operator \eqref{opp1p1} for $\xi=1$  at $\hbar={3\pi \over 2} \re^{\ri \pi /4}$ obtained by solving \eqref{EQC} and by using  \eqref{WvsE}. We denote by $n$ the order at which we truncate the $Q_1$-series in \eqref{reftv}. The numerical value was computed by using matrices of size 300.} \label{hTaBb}
        \end{center}
   \end{table}

When $\hbar \in \IR_+$, the non-perturbative corrections to the all-orders WKB quantization condition of \cite{ns} 
are crucial to eliminate poles and to even write down a sensible 
answer, as first pointed out in \cite{km}. However, when $\hbar$ is complex, there are no poles to start with, 
and the all-orders WKB quantization condition (given by the first term in the l.h.s. of (\ref{EQC})) makes sense. However, the non-perturbative corrections are 
still required to obtain the right spectrum. In \tabref{TB2} we show the ground state energy obtained by neglecting non-perturbative effects  in \eqref{EQC}, i.e. by neglecting the term 
\be \label{plus} {\partial \over \partial t_j} F^{\rm inst}_{\rm NS}\left({2 \pi \over \hbar} {\boldsymbol{t}}(\hbar), {4 \pi^2  \over \hbar} \right), \ee
and computed for a particular complex value of $\hbar$. As we increase the number of terms in the $Q_1$ series, the putative ground state energy converges, albeit to a 
wrong value. In view of this, we conclude the following: 
\begin{enumerate}
\item  If $\hbar$ is real, the non--perturbative terms \eqref{plus} cancel the poles appearing in the all-orders WKB quantization condition, and lead to a convergent expansion for the l.h.s. of (\ref{gns}).
\item  If $\hbar $ is not real, the l.h.s.~of \eqref{gns}, after the resummation implemented by the gauge theory instanton calculation, has good convergence properties, even without the non--perturbative terms \eqref{plus}. Nevertheless, if we do not include the non--perturbative corrections, the quantization condition does not lead to the correct energy levels of the operator. 
\end{enumerate}

\begin{table}  [H]  \begin{center}
 \begin{tabular}{ccccc}
	
	$n$ & $E_0 ^{\rm NS}$    \\ \hline
	3  &2.1233154521796 + 0.2381282827804  \ri\\ 
	 5 &2.1233157637682 + 0.2381283851298 \ri \\ 
	 6 &2.1233157636916 + 0.2381283852383 \ri \\ \hline
Num     &2.1234991968312 + 0.2382001183808 \ri           &
	     \end{tabular}
        \caption{ The first energy level  at $\hbar=3+\ri$ obtained from \eqref{EQC} by neglecting non-perturbative effects, i.e. the term \eqref{plus}. We denote by $n$ the order at which we truncate the $Q_1$-series in \eqref{reftv}. The numerical value is computed  by using matrices of size 300. This shows how the quantization condition 
        converges to the wrong value if we do not include  the non--perturbative effects. }
      \label{TB2}
        \end{center}
   \end{table}

In \cite{kaserg}, Kashaev and Sergeev have studied the spectral problem of the operator (\ref{opp1p1}) for $\xi=1$ and complex values of $\hbar$ of the form 
\be
\hbar = 2 \pi \re^{2\ri \theta}, \qquad \theta \in [0, \pi/2). 
\ee
In particular, they have obtained exact quantization conditions for the spectrum, which they have studied numerically for $\hbar=2 \pi \ri$.    
We have checked that the results obtained with our proposal agree 
with their results. It turns out that, when $\hbar=\pm 2 \pi \ri$ and the quantum number $n$ is of the form
   \be 
   n = 2m (m+1), \quad m\in \IN 
   \ee
   the quantization condition  \eqref{EQC} becomes simply
   \be  \label{tco} \ba  t_1^{(n)}(2 \pi \ri,1)= &{\pm }(1+\ri)(1+2m)\pi,  \\
   t_1^{(n)}(-2 \pi \ri,1)= &{\pm }(1-\ri)(1+2m)\pi.
   \ea\ee
   In particular, for these values of the K\"ahler parameter \eqref{tco}, the series in $Q_i$ and the one in $Q_i^D$  inside \eqref{EQC} mutually cancel. When $m=0$, 
by using  \eqref{tco} and \eqref{WvsE}, we have
\be\label{E0x} E_0(\pm 2 \pi \ri )=1.5248292052302074168 \cdots \pm 1.5707963267948966192 \ri \cdots 
\ee
which agrees with the numerical computation and with the result of \cite{kaserg}.

 \section{The quantum Mathieu operator}\label{s4}
 
 So far, topological string theory has been used to solve spectral problems that arise by taking a ``real" slice of the mirror curve and applying Weyl quantisation to the mirror curve, such as for instance \eqref{opp1p1}. However, as already pointed out in \cite{ns} in the case of spectral problems arising from 4d gauge theories, one can 
 obtain other, related spectral problems by considering ``imaginary" slices of the variables $x$, $p$. This involves the following transformation of the operators 
 $\mathsf x, \mathsf p$ introduced in \eqref{quant}:
 \be \mathsf x \to \ri \mathsf x, \quad  \mathsf p \to \ri \mathsf p. \ee
 Such transformations change completely the nature of the spectral problem. For instance, the rotation of $\mathsf x$ leads to the analogue of a periodic 
 potential. In this section we study the operator obtained from \eqref{opp1p1} after such a rotation, i.e. we study de spectrum of \footnote{We have  shifted $\mathsf x$ by ${1\over 2}\log \xi $ w.r.t.  \eqref{opp1p1}.}
\be 
\label{hapfp1p1} {\mathsf O}_{\text{qM}}= R(\re^{\ri \mathsf x}+ \re^{- \ri \mathsf x})+\re^{\mathsf p}+ \re^{- \mathsf p} , \quad [\mathsf x, \mathsf p]=\ri \hbar, 
\ee
where $R=\xi^{1/2}$. This is a quantum deformation of the standard Mathieu operator 
\be
\mO_{\text{M}}= \mm^2 + 2 R \cos(\mx), 
\ee
and we will call it the {\it quantum Mathieu operator}.  The eigenvalue equation for this operator 
\be
\mO_{\text{qM}} \psi(x) = \epsilon \psi(x)
\ee
can be also written as
\be \label{myeq} 
\psi(x+ \ri \hbar) + \psi(x- \ri \hbar) +V(x) \psi(x)= \epsilon \psi(x), 
\ee
where 
\be
\label{potper}
 V(x)= 2 R \cos(x)
 \ee
 is the potential appearing in the standard Mathieu equation
 \be \label{mat}- \hbar ^2 {\rd^2 \psi(x)\over \rd x^2} +2R \cos(x)\psi(x)= u \psi(x). 
 \ee
Since (\ref{myeq}) involves a periodic potential $V(x)$, with period $a= 2 \pi$, 
we will have a band structure as for the standard Mathieu equation (\ref{mat}). 
In this section we first solve the spectral problem for \eqref{hapfp1p1} by using standard techniques for periodic potentials, 
and then we will express the result in terms of topological string quantities\footnote{The quantum Mathieu operator has been also studied in \cite{krefl3}. 
However, the analysis in \cite{krefl3} seems to be restricted to a particular value of the quasimomentum, and is based on an asymptotic WKB expansion.}. As we will 
see, the spectrum is determined exactly  
by the quantum mirror map. In particular, 
no non-perturbative corrections seem to be needed in this case. We note that the Mathieu and modified Mathieu equation, when solved in terms of gauge theory, 
involve the quantum A and B periods of the Seiberg--Witten curve, respectively \cite{ns}. The solution of the quantum Mathieu case is similar, since the quantum mirror map 
can be regarded as the quantum A-period of the mirror curve. 

\subsection{The spectrum from the central equation}

\begin{figure}[tb]
\begin{center}
\begin{tabular}{cc}
\resizebox{70mm}{!}{\includegraphics{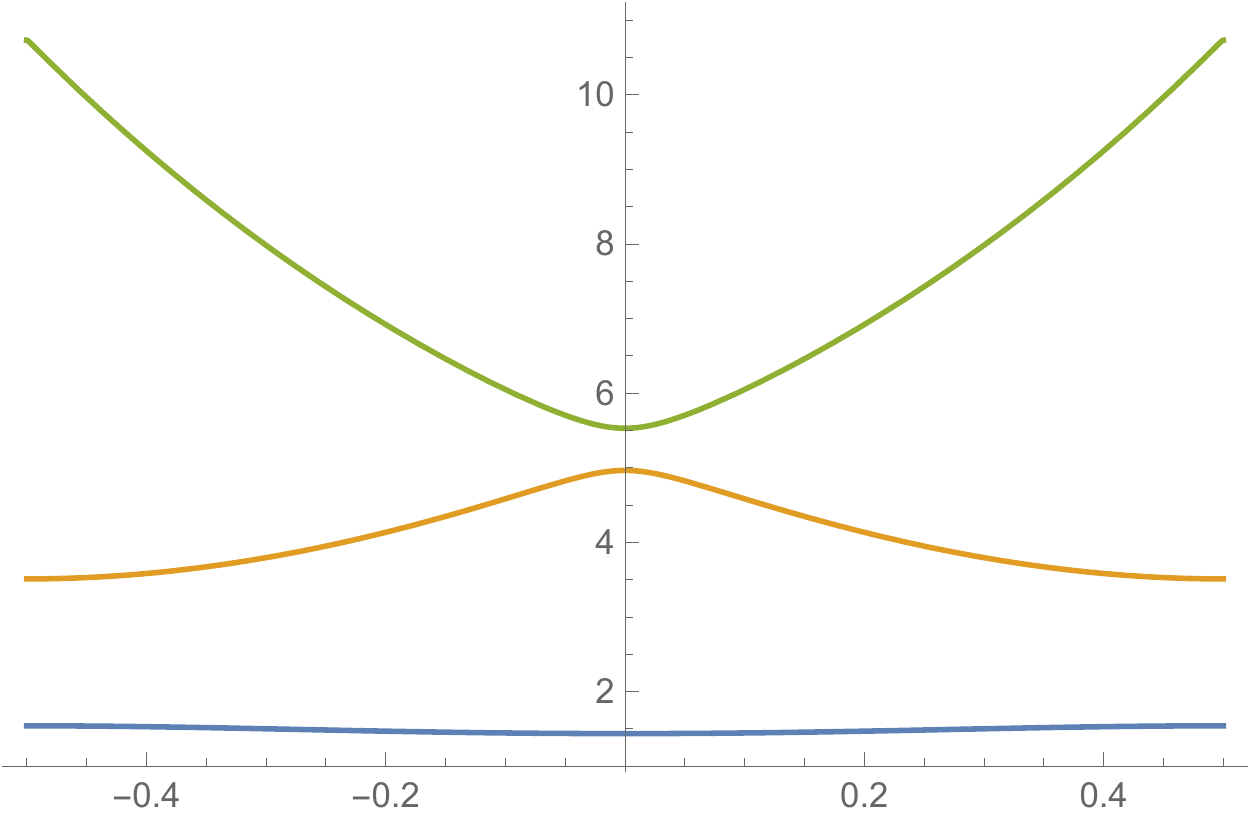}}
\hspace{3mm}
&
\resizebox{70mm}{!}{\includegraphics{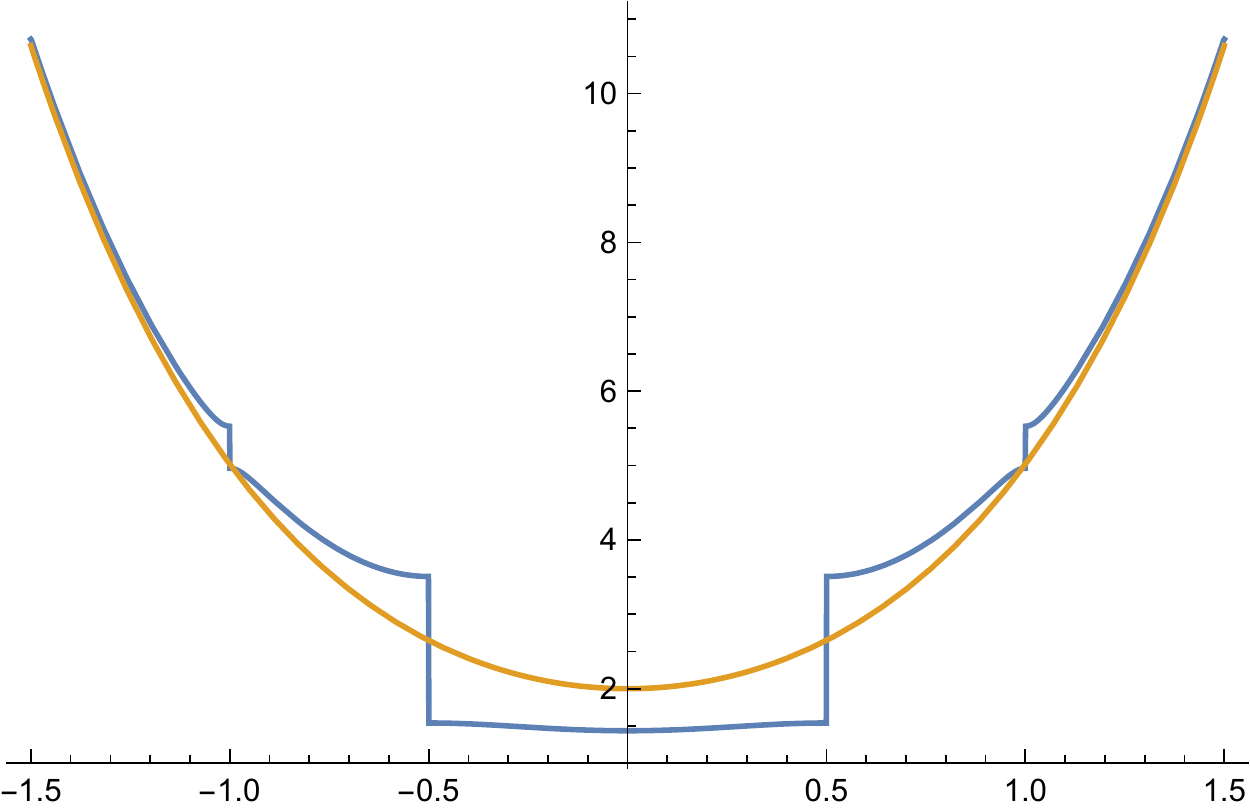}}
\end{tabular}
\end{center}
  \caption{The first three energy bands  of the quantum Mathieu equation for $R=1$, $\hbar =\pi/2$, represented in the first Brillouin zone (left) and 
  in the extended zone (right), where we also plot the dispersion relation $2 \cosh(\hbar k)$ for the free theory. 
}
\label{qu-math-fig}
\end{figure}

A standard tool to solve a periodic potential (see for example \cite{am}, Chapter 8) consists of writing a Fourier expansion for the Bloch wavefunction 
and solve for its coefficients. This leads to the so-called {\it central equation}, which can be used to calculate the spectrum numerically and, sometimes, analytically. 
Let us then write down the Bloch wavefunction
\be
\label{bloch-fourier}
\psi(x)= \sum_q c_q \re^{\ri q x}, 
\ee
where $q$ is of the form 
\be
q= k + {2 \pi \over a} \ell, \qquad \ell \in \IZ, \quad k \in \left[ -{\pi \over a}, {\pi \over a}\right],
\ee
$k$ is the quasimomentum, and $a$ is the period of the potential. We will denote the corresponding coefficient by 
\be
c_q= c_\ell^{(k)}.
\ee
For a general periodic potential, one uses the Fourier decomposition
\be
V(x) = \sum_{r \in \IZ} V_r \re^{2 \pi \ri r x/a}. 
\ee
In our case, $a=2 \pi$ and $V_r=0$ unless $r=\pm1$. We have in addition a non-standard kinetic term $2 \cosh(\mm)$. 
By plugging (\ref{bloch-fourier}) inside \eqref{myeq}, we find a set of algebraic equations for the coefficients $c_q$:
\be
\label{central-def}
2 \cosh \left(\hbar q \right) c_q +  R \left( c_{q-1}+ c_{q+1} \right)=\epsilon(k, \hbar, R) c_q.
\ee
This is the analogue of the central equation for the quantum Mathieu operator. By using the explicit expression for $q$, we can also write it as
\be
\label{central-cosh}
2 \cosh \hbar ( k +\ell) c_\ell^{(k)} +  R \left( c^{(k)}_{\ell-1}+ c^{(k)}_{\ell+1} \right)=\epsilon(k, \hbar, R) c_\ell^{(k)}. 
\ee
For a fixed $k$, this gives an infinite dimensional matrix equation for $c_\ell^{(k)}$. In practice, one truncates the equation for a given $\ell_{\rm max} \ge |\ell |$, to obtain $2\ell_{\rm max}+1$ 
values of $\epsilon(k, \hbar, R)$.
These are the energy bands, as a function of the quasi-momentum in the Brillouin zone
\be k \in [-1/2, 1/2]. \ee
For values of $\hbar, \, R \sim 1$,  truncating this matrix equation to finite size gives extremely efficient results for the 
energy bands. Even $\ell_{\rm max}=5$ gives more than 100 correct digits for the ground state energy. In \figref{qu-math-fig} we show the first three energy 
bands for $\hbar=\pi/2$, $R=1$.

The central equation can be also used to solve for the band spectrum analytically, in terms of a power series. In the case of the Mathieu equation, 
this is nicely explained in section 2 of \cite{slater}. This analysis can be easily extended to the quantum Mathieu operator. There are two different 
cases, depending on whether the quasi-momentum is in the interior or at the edges of the bands. If $k$ is an interior value, we can write the following 
ansatz for the band energy, 
\be \label{epsan}
\epsilon(k, \hbar, R) = 2 \cosh(k \hbar) +\sum_{\ell \ge 1} \epsilon_\ell(k, \hbar) R^{2\ell},  
\ee
where the first term is the dispersion relation for the free theory. We will now assume the normalization condition that $c_{k}=1$ for a fixed $k$, and that 
\be \label{cond}
c_{k\pm  n}= R^n \sum_{\ell\ge 0} c^{(\ell)}_{\pm n}(k)R^{2\ell}. 
\ee
The central equation for $k\pm 1$ and at leading order in $R$ gives
\be
 \left( 2 \cosh (k\pm 1) \hbar - 2 \cosh(k \hbar) \right) R c^{(0)}_{\pm 1}(k) + R + \CO(R^2)=1, 
 \ee
 so we find 
 \be
 c^{(0)}_{\pm 1}(k)= -{1\over 2 \cosh (k\pm 1) \hbar - 2 \cosh(k \hbar)}. 
 \ee
 In addition, by using the central equation for $k$, we find at leading order 
 \be\label{epssol} 
 \epsilon_1(k, \hbar)=  c^{(0)}_{1}(k)+  c^{(0)}_{-1}(k)= {( Q_2^{1/2}+ Q_2^{-1/2}) Q_1 { R^{-2}} \over \left(q^{1/2} Q_2- q^{-1/2}\right)\left( q^{-1/2} Q_2 - q^{1/2} \right)}, 
 \ee
 with
 \be Q_1=R^2 \re^{2k \hbar} , \quad Q_2=\re^{2 k \hbar}, \quad q=\re^{-\hbar}. \ee
 In particular, \eqref{epssol} reproduces precisely the first term in the expansion of $W(Q_1, Q_2, q)$ as computed by instanton calculus in section \ref{s3}. 
 It can be easily checked that this is also the case for the next few higher order terms.  
  \begin{table}  \begin{center}
    \begin{tabular}{cc}
	 order & $ \epsilon_0(0, 2\pi, 1) $   \\ \hline
		1 &   \underline{1.9962511}25995166296078333  \\ 
	 3 &  \underline{1.9962511523137339}27269664 \\
	 5 &   \underline{1.996251152313733933761514}\\   \hline
  Num  &       1.996251152313733933761514  \\   \hline
	     \end{tabular}
        \caption{ The energy $\epsilon(0, 2\pi, 1)$ for the quantum Mathieu equation at the center of the band, as computed from (\ref{per-eqc}). The order denotes the number of terms retained in the expansion of $W(Q_1, Q_2, q)$ in powers of  $Q_1$.}          \label{eo-2pi}   \end{center}
   \end{table}
We conjecture that this is true to all orders in the series expansion (\ref{epsan}), so that
\be
\label{per-eqc}
\epsilon(k, \hbar, R)= W \left( R^2 \re^{2k \hbar},\re^{2 k \hbar},  \re^{-\hbar} \right). 
\ee
We can then use the quantum mirror map to compute the band energies. An example of such a calculation is shown in table \ref{eo-2pi}.
We also observe that the solution \eqref{per-eqc} gives the correct band energies for complex values of $\hbar$ as far as ${\rm Re}(\hbar) \neq 0$\footnote{When $\hbar$ is purely imaginary, 
the spectral problem of the quantum Mathieu operator might be related to the one for the fully periodic operator (\ref{harper}).} .


 The equation (\ref{per-eqc}) gives an analytic expression for the band energies, in terms of a series expansion for $W(Q_1, Q_2, q)$ to obtain 
 concrete results. This series is convergent, so we do not expect to have non-perturbative corrections to (\ref{per-eqc}). It has however a finite radius of convergence. 
 Let us consider for example the center of the first band $k=0$, where we expect the maximal 
 difference w.r.t. the free theory. We have the following structure, 
 \be
  \epsilon_\ell(0, \hbar)= {f_\ell(q) \over (q-1)^{4 \ell-2}}, 
  \ee
  where $f_\ell(q)$ is regular at $q=1$. In addition, numerical experiments indicate that,
  \be
  f_{\ell}(q) \sim (-1)^{\ell} (f(q))^{\ell}, \qquad \ell \gg 1, 
  \ee
  where $f(q)>0$ whenever $q\in (0,1]$. This means that the coefficients $  \epsilon_\ell(0, \hbar)$ grow, for $\ell$ large, as 
  \be \label{divep}
  \epsilon_\ell(0, \hbar) \sim \left[ -{f(q)  \over (q-1)^4} \right]^\ell, 
  \ee
  so the radius of convergence of this series is approximately given by 
  \be \label{rc}
  R^2 \sim {   (q-1)^4 \over  f(q) }. 
  \ee
 This goes to zero as $\hbar \to 0$, so we don't expect convergence for small values of $\hbar$, as we indeed observe numerically by computing the coefficients of \eqref{epsan} to high order. Nevertheless, since the series is alternating, it is expected that it can be resummed to a convergent function on the positive real axis of $R$.  In particular, 
 the convergence of (\ref{epsan}) as $\hbar \to 0$ can be remarkably improved by using standard algorithms such as Pad\'e approximants or Shanks transformations (see 
 for example \cite{bender-book}). Let us illustrate this in one example, namely  $k=0$ and $\hbar= \pi/3$. 
It is useful to define the truncated series
\be\label{mye} \epsilon^{(N)}[R]= \sum_{n\geq 0}^N \epsilon_{n}(0, {\pi / 3})R^{2n} \ee
and its  Shanks transformation\footnote{In this example, the Shanks transform turns out to be a bit more efficient than the Pad\'e approximant.}  as 
\be S( N,R)={ \epsilon^{(N +1)}[R] \epsilon^{(N -1)}[R]-  (\epsilon^{(N)}[R])^2 \over  \epsilon^{(N+1)}[R]-2 \epsilon^{(N)}[R]+ \epsilon^{(N-1)}[R]}.\ee
Likewise we denote by
\be S_n( N,R), \quad N>n\ee
the quantity obtained after applying $n$ times the Shanks transformation. For instance
\be   S_{2}( N,R) ={ S({N +1} ,R)S(N -1,R)-  (S(N,R))^2 \over S({N +1} ,R)-2 S({N } ,R)+ S({N -1} ,R)}.\ee
For $R=1$ \eqref{mye}  diverges at large $N$  as expected from \eqref{rc} and shown in the first column of \tabref{tb1}. Nevertheless by applying Shanks transformation repeatedly we can improve the convergence remarkably as shown for instance in  \tabref{tb1}.
 
\begin{table}
\begin{center}
 \begin{tabular}{ccccc}
	$N$ & $ \epsilon^{(N)}[1]$ &$S_3( N+3,1) $ & $S_6( N+6,1) $ &$S_9( N+9,1) $    \\ \hline
	1 &0.334  & 1.0228922 &1.0201454 & 1.02014601733678\\ 
        2 &2.423  & 1.0187015 &1.0201458&1.02014600931881\\ 
         3&-3.038 & 1.0212857 &1.0201461&1.02014600851393 \\ 
	4&14.864 & 1.0188895 &1.0201458 &1.02014600944477\\ 
	5& -50.900&1.0219159 &1.0201462 &1.02014600942947 \\ \hline 
  Num &   &  & & 1.02014600942838
	     \end{tabular}
        \caption{The truncated (divergent) series \eqref{mye} for $k=0$, $\hbar=\pi/3$, $R=1$  together with its  $3^{\rm rd}$, $6^{\rm th}$ and  $9^{\rm th}$  Shanks transform.  In the last line the numerical value obtained by truncating the matrix equation is given.
        }  \label{tb1}
       \end{center}
   \end{table}
   
The above solution for the energies (\ref{per-eqc}) is valid in the interior of the band. As in the standard Mathieu equation, there are poles at the 
edges of the bands, i.e. when 
$k=\pm n/2$, $n\in \IZ_{>0}$ (the poles at $k= \pm 1/2$ are manifest in (\ref{epssol})). This case has then to be treated separately, and we can follow again the methods of \cite{slater}. 
To solve the central equation at the edges we have to change the ansatz \eqref{cond}, \eqref{epsan}. 
We fix $k={1/2}$ at the edge of the Brillouin zone and we consider the following ansatz
 \be \label{epsan2}
\epsilon(\alpha, {1/ 2}, \hbar, R) = 2 \cosh( \hbar /2) +\sum_{\ell \ge 1} \epsilon_\ell(\alpha,{1/ 2}, \hbar) R^{\ell},  
\ee
with
\be c_{1/2}=c^{(1/2)}_0=1, \quad c^{(1/2)}_{-1}=\alpha, \quad \alpha^2=1\ee
\be\label{anc}  c^{(1/2)}_{ n} =R^n \sum_{\ell\geq 0} d^{\ell}_{ n} R^{\ell}, \quad n>0.\ee
By solving \eqref{central-cosh} with this ansatz we find
 \be \label{ep2}  \ba & \epsilon\left(\alpha,1/2, \hbar, R\right)=2 \cosh \left(\frac{\hbar }{2}\right)+\alpha  R +\frac{R^2}{2 \cosh \left(\frac{\hbar }{2}\right)-2 \cosh \left(\frac{3 \hbar }{2}\right)}-\frac{\alpha  R^3}{4 \left(\cosh \left(\frac{\hbar }{2}\right)-\cosh \left(\frac{3 \hbar }{2}\right)\right)^2}\\
 &-\frac{R^4 \text{csch}^3\left(\frac{\hbar }{2}\right) \text{csch}^3(\hbar )}{128 \cosh (\hbar )+64}+\frac{\alpha R^5 (4 \cosh (\hbar )+4 \cosh (2 \hbar )+3) \coth ^2\left(\frac{\hbar }{2}\right) \text{csch}^6(\hbar )}{64 (2 \cosh (\hbar )+1)^2}+\mathcal{O}(R^6).
   \ea \ee
  By setting $\alpha=-1$ we obtain the energy at the edge of the first band, while $\alpha=1$ gives the energy at the edge of the second band. 
   Some results are shown in \tabref{ep1} and \tabref{ep11} and are in perfect agreement with the numerical computations. 
    \begin{table} [h!]\begin{center}  
    \begin{tabular}{cc} 
	 order & $  \epsilon\left(-1,{1\over 2}, 2\pi , 1\right) $   \\ \hline
	 1 &  \underline{22.183}9065510430412555035041051  \\ 
	 3 &  \underline{22.18382570679668}48185142066748   \\ 
	 5 &  \underline{22.183825706796683748}0949560885 \\
	 7 &  \underline{22.183825706796683748105320697}7\\   \hline
  Num  &                    22.1838257067966837481053206975    \\   \hline
	     \end{tabular} 
 \caption{ The energy $  \epsilon\left(-1,{1\over 2}, 2\pi , 1\right)$ for the quantum Mathieu equation at the edge of the first band, as computed from (\ref{ep2}). The order denotes the number of terms retained in the expansion of $ \epsilon\left(-1,{1\over 2}, 2\pi , R\right)$ in powers of  $R$.}           \label{ep1}     \end{center}
 \end{table}  
       \begin{table} [h!] \begin{center}
    \begin{tabular}{cc}
	 order & $  \epsilon\left(1,{1\over 2}, 2\pi , 1\right) $   \\ \hline
	 1 &  \underline{24.183}90655104304125550350410512  \\ 
	 3 &  \underline{24.18382569372298}652860218359466  \\ 
	 5 &  \underline{24.1838256937229856287}8563284976 \\
	 7 &  \underline{24.183825693722985628795996908}71\\   \hline
  Num  &                   24.18382569372298562879599690846                 \\   \hline
	     \end{tabular}          \caption{ The energy $  \epsilon\left(1,{1\over 2}, 2\pi , 1\right)$ for the quantum Mathieu equation at the edge of the second band, as computed from (\ref{ep2}). The order denotes the number of terms retained in the expansion of $ \epsilon\left(1,{1\over 2}, 2\pi , R\right)$ in powers of  $R$.}      \label{ep11}
   \end{center}
   \end{table}
Notice that for $k=-{1\over 2}<0$ one has to replace the ansatz \eqref{anc} by
\be c_{1/2}=c^{(1/2)}_0=1, \quad c^{(1/2)}_{1}=\alpha\ee
\be  c^{(1/2)}_{ -n} =R^n \sum_{\ell\geq 0} d^{\ell}_{ n} R^{\ell}, \quad n>0.\ee
However, at the end the result is the same, namely $\epsilon\left(\alpha,{1\over 2},\hbar , R\right)=\epsilon\left(\alpha,-{1\over 2},\hbar , R\right)$, as expected.
We note that the solution \eqref{ep2}  also holds for complex values of $\hbar$ as far as $\rm{Re} (\hbar) \neq 0$. Other values of $k$, corresponding to other edges of the bands, can be 
worked out similarly. 

The series \eqref{epsan2} is convergent with a finite radius of convergence, so we do not expect to have non-perturbative corrections. Numerically we observe a behaviour similar to \eqref{divep}, with a zero radius of convergence as $\hbar \rightarrow 0$. As before, we can improve the convergence  by using standard  algorithmss. For the series \eqref{epsan2}, 
the Pad\'e approximant turns out to be well suited. An example is shown on Table \ref{tbp}. 
We use the same convention as in \cite{gmz} for the Pad\'e approximant.
\begin{table} [h!]
\begin{center} 
 \begin{tabular}{ccccc}
	   $n$  & $P_n^{\epsilon}(1)$ \\ \hline
	10 & \underline{0.5}156117750532916   \\
	20 & \underline{0.52058}03657529506  \\
	30 & \underline{0.5205897}527191174 \\
	40 & \underline{0.52058973}46606964\\
	50 & \underline{0.520589735365}0857 \\
        55 &                 \underline{0.52058973536513}56 \\ \hline 
  Num &                    0.5205897353651301	   \\
     \end{tabular} 
        \caption{ The Pad\'e approximant of order $n$  of for the series \eqref{epsan2} with $k=1/2$ , $\hbar={\pi/6}$, $R=1$. We computed  the coefficients in the series expansion \eqref{epsan2} up to $R^{56}$.         }  \label{tbp}
       \end{center}
   \end{table}

Finally, we note that the even powers of $R$ in \eqref{epsan2} can be obtained from the Wilson loop 
as
\be \sum_{n\geq 0}  \epsilon_{2n}(\alpha, 1/2, \hbar) R^{2n}= {\rm{Res}}_{k=1/2} \left[W(\re^{2 k \hbar} R^2, \re^{2 k \hbar}  ,\re^{- \hbar}) (k-1/2)^{-1}\right].\ee
It would be interesting to see whether there is a similar interpretation for the odd powers of $R$ in \eqref{epsan2}.

\subsection{Four dimensional limit}
In order to recover the  Mathieu equation (\ref{mat}) from the quantum Mathieu equation \eqref{myeq} 
one has to implement the standard four dimensional limit, i.e. we replace
 \be \label{4dlim}
\hbar\to {  \beta \hbar}, \quad R\to \beta^2 R, 
\ee
and we take the limit $\beta \rightarrow 0$. 
 Then, away from the edges of the bands, we obtain the well-known result expressing the band energies for the Mathieu equation 
 in terms of the NS free energy of the $SU(2)$, $\CN=2$ theory, namely (see for example \cite{he-miao,kpt,basar-dunne,du-mathieu})
\be 
\label{userie} \ba  u(k, R, \hbar)&=k^2 \hbar^2- \Lambda \partial_{\Lambda}F^{\rm NS}(\Lambda, k \hbar,  \ri \hbar) \ea\ee
where we set $ \Lambda=R^2$, and 
\be F^{\rm NS}(\Lambda, a , \hbar)=-\frac{2 \Lambda }{4 a^2+\hbar ^2}+\frac{\Lambda ^2 \left(7 \hbar ^2-20 a^2\right)}{4 \left(a^2+\hbar ^2\right) \left(4 a^2+\hbar ^2\right)^3}+\mathcal{O}(\Lambda^4).\ee
These results can be easily obtained by solving the central equation for \eqref{mat} instead of taking the four dimensional limit \eqref{4dlim}. More precisely, 
the central equation for \eqref{mat} reads
\be
\label{central-cos}
\hbar^2 \left( k + {2 \pi \over a} \ell\right)^2 c_\ell^{(k)} +  R \left( c^{(k)}_{\ell-1}+ c^{(k)}_{\ell+1} \right)=u(k,R,\hbar) c_\ell^{(k)}. 
\ee
By solving \eqref{central-cos} at the edge of the first zone, i.e. for $k=1/2$, one finds \cite{slater, mattab}
\be   \label{userie2}u(1/2, R, \hbar)=u_o(\alpha,1/2,R,\hbar)+u_e(1/2,R,\hbar),
 \ee
where $u_{e,o}$ are the even and odd powers of $R$, respectively. They have the explicit expression, 
\be \ba   u_o(\alpha,1/2,R, \hbar)&= R\alpha-\alpha\frac{R^3}{4 \hbar^4}  +\alpha\frac{11 R^5}{144 \hbar^8} +\alpha \frac{55 R^7}{2304 \hbar^{12}}+\cdots\\
u_e(1/2,R, \hbar)&={1\over 4}\hbar^2 -\frac{R^2}{2 \hbar^2} -\frac{R^4}{24  \hbar^6}+ \frac{49 R^6}{576  \hbar^{10}}+\cdots
\ea\ee
 By choosing $\alpha=-1$ one obtains the energy at the edge of the first band, while $\alpha=1$  gives the energy at the edge of the second band.
As in the quantum Mathieu case, the even powers are still determined by the NS free energy. Indeed, one has
\be 
u_e(1/2,R, \hbar)={\rm Res}_{k=1/2}\left({u(k, R, \hbar )\over k-1/2 }\right)=
{\rm Res}_{ k=1/2}\left({k^2 \hbar^2- \Lambda \partial_{\Lambda}F^{\rm NS}(\Lambda, k \hbar ,  \ri \hbar) \over k-1/2 }\right) \biggl|_{\Lambda=R^2}.\ee
Again, it would be interesting to find an expression for the odd powers $u_o(\alpha,1/2, R, \hbar)$ in terms of gauge theory.

As a final observation, we note that, as in the case of quantum Mathieu equation, the series \eqref{userie2}, \eqref{userie} have a finite radius 
of convergence which goes to zero at $\hbar=0$. Nevertheless, one can also use Pad\'e approximants to calculate the band energies. 

\section{Conclusions}

In this paper we have taken the first steps to extend the TS/ST correspondence, as formulated in \cite{ghm,cgm}, to complex values of the Planck constant. 
We have noted that, at least when the toric CY engineers a 5d gauge theory, one can resum the series appearing in the exact quantization 
condition, in the form given in \cite{wzh}. The results obtained in this way are in agreement with numerical calculations of the spectrum. One interesting 
outcome of our computations is that the non-perturbative effects detected in \cite{ghm,cgm} are crucial to obtain the correct results, even in 
situations where the all-orders WKB quantization condition of \cite{ns} makes sense. 

Clearly, there are many problems that one should address in order to have a full understanding of the complex side of the 
TS/ST correspondence. First of all, our results cover a small subset of all toric geometries, namely those directly related to gauge theories. 
For other geometries, like local $\IP^2$, it is not clear how to resum the (refined) topological free energies, the grand potential and the quantum mirror map, 
in order to obtain convergent expansions. We should also note that, when $\hbar$ is real, the correspondence allows one to calculate the spectral traces and the spectral determinant of the operators. 
It would be interesting to see how this can be achieved in the complex case, even in examples related to gauge theories. 

In this paper we have also considered other slices in the quantization of the mirror cuve, focusing for simplicity in the quantum Mathieu operator associated to 
local $\IF_0$. The solution for the resulting band structure only involves the quantum mirror map and is much simpler than the solution of the trace class operators 
considered in \cite{ghm}. 
The underlying reason, as pointed out in \cite{basar-dunne}, is that we can regard the full periodic potential as a perturbation of a free particle on a circle. 
This leads to the convergent expansion (\ref{epsan}), which is nothing but the quantum mirror map. Non-perturbative effects are not required to compute the 
energies of the bands, in contrast to what happens in the quantization scheme of \cite{ghm}, where these effects are crucial. Since 
the expansions break down at the edges, one can however consider the energy 
splitting of the bands as some sort of non-perturbative effect, as advocated in \cite{avron-simon,basar-dunne,du-mathieu} for the standard Mathieu equation. 
It would be interesting to study the splitting for the quantum Mathieu equation from that point of view. 

The analysis of the operators along ``imaginary" slices should be extended to more general geometries. One could also study operators obtained 
by going to the imaginary slices for both $\mx$ and $\mm$. In the case of local $\IF_0$, this gives the ``fully periodic" operator
\be
\label{harper}
\mO=2 \cos(\mm)+  2 R \cos(\mx). 
\ee
It is easy to see that this is equivalent to the Harper operator, leading to the famous Hofstadter butterfly. It has been observed 
that the quantum geometry of local $\IF_0$ is closely related to some aspects of this problem \cite{butterfly}. It would be interesting 
to see whether topological string theory sheds further light on the subtle spectral properties of the Harper operator. 

\section*{Acknowledgements}
We would like to thank S. Codesido, S. Zakany and specially R. Kashaev for useful discussions. 
The work of M.M. is supported in part by the Fonds National Suisse, 
subsidies 200021-156995 and 200020-141329, and by the NCCR 51NF40-141869 ``The Mathematics of Physics" (SwissMAP).  A.G. acknowledges support by INFN Iniziativa Specifica ST$\&$FI.

\appendix
\section{The grand potential} \label{defJ}

In this section we review the definition of the topological string grand potential.
For this we first need to introduce various quantities appearing in topological string theory and in quantum geometry.  We follow \cite{cgm,mmrev}, where more 
details and references can be found. 

Let $X$ be a toric CY geometry as in section \ref{s2}. It is convenient to write the ``true" complex moduli  \eqref{tm} of $X$ as
\be \kappa_i=\re^{\mu_i},
\ee
and to introduce the Batyrev coordinates $z_i$ such that
\be
\label{zmu}
-\log \, z_i= \sum_{j=1}^{g_\Sigma}C_{ij} \mu_j + \sum_{k=1}^{r_\Sigma} \alpha_{ik}\log {\xi_k}, \qquad i=1, \cdots, n_\Sigma,
\ee
where $C_{ij}$ is a $ n_{\Sigma}\times g_{\Sigma} $ matrix which is determined by the toric data of $X$ as explained in \cite{kpsw}. 
The Kahler  moduli $t_i$  and the Batyrev coordinates are related by the mirror map
\be \label{mir} -t_i=\log z_i+ \Pi_i( \bf z), \ee
where $\Pi_i$ is a power series in $z_j$. We also denote $Q_i=\re^{-t_i}.$ As explained in \cite{acdkv}, we can promote  \eqref{mir} to a quantum mirror map
\be\label{qmir} -t_i(\hbar)=\log z_i+ \Pi_i( \bf z, \hbar). 
\ee
We introduce the topological string free energy
\be F^{\rm top}\left({\bf t}, g_s\right)={1\over 6 g_s^2} a_{ijk} t_i t_j t_k +b_i t_i +F^{\rm GV}\left({\bf t}, g_s\right)\ee
where
\be
\label{GVgf}
F^{\rm GV}\left({\bf t}, g_s\right)=\sum_{g\ge 0} \sum_{\bf d} \sum_{w=1}^\infty {1\over w} n_g^{ {\bf d}} \left(2 \sin { w g_s \over 2} \right)^{2g-2} \re^{-w {\bf d} \cdot {\bf t}}
\ee
is written in terms of the Gopakumar--Vafa (GV) invariants $n_g^{ {\bf d}}$ of the underlying geometry $X$. 
Similarly, the Nekrasov--Shatahsvili (NS) free energy of $X$ is defined as
\be
\label{NS-j}
F^{\rm NS}({\bf t}, \hbar) ={1\over 6 \hbar} a_{ijk} t_i t_j t_k +b^{\rm NS}_i t_i \hbar +\sum_{j_L, j_R} \sum_{w, {\bf d} } 
N^{{\bf d}}_{j_L, j_R}  \frac{\sin\frac{\hbar w}{2}(2j_L+1)\sin\frac{\hbar w}{2}(2j_R+1)}{2 w^2 \sin^3\frac{\hbar w}{2}} \re^{-w {\bf d}\cdot{\bf  t}},
\ee
where $N^{{\bf d}}_{j_L, j_R}$ are refined BPS invariants. We also need a B-field satisfying 
\be
\label{B-prop}
(-1)^{2j_L + 2 j_R+1}= (-1)^{{\bf B} \cdot {\bf d}},
\ee
for the indices labelling non-vanishing invariants $N^{{\bf d}}_{j_L, j_R}$.
  
The topological string grand potential is defined as
\be
\label{jtotal}
\mathsf{J}_{X}(\boldsymbol{\mu}, \boldsymbol{\xi},\hbar) = \mathsf{J}^{\rm WKB}_X (\boldsymbol{\mu}, \boldsymbol{\xi},\hbar)+ \mathsf{J}^{\rm WS}_X 
(\boldsymbol{\mu},  \boldsymbol{\xi} , \hbar), 
\ee
where
\be
\label{jws}
\mathsf{J}^{\rm WS}_X(\boldsymbol{\mu}, \boldsymbol{\xi}, \hbar)=F^{\rm GV}\left( {2 \pi \over \hbar}{\bf t}(\hbar)+ \pi \ri {\bf B} , {4 \pi^2 \over \hbar} \right),
\ee
and 
\be
\label{jm2}
\mathsf{J}^{\rm WKB}_X(\boldsymbol{\mu}, \boldsymbol{\xi}, \hbar)= {t_i(\hbar) \over 2 \pi}   {\partial F^{\rm NS}({\bf t}(\hbar), \hbar) \over \partial t_i} 
+{\hbar^2 \over 2 \pi} {\partial \over \partial \hbar} \left(  {F^{\rm NS}({\bf t}(\hbar), \hbar) \over \hbar} \right) + {2 \pi \over \hbar} b_i t_i(\hbar) + A({\boldsymbol \xi}, \hbar). 
\ee
Here, $ A({\boldsymbol \xi}, \hbar)$ is a moduli-independent contribution which can be related to the contribution of constant maps in topological string theory. 

We will also use the following notation
 \be \label{NS-inst}
F^{\rm NS,inst}({\bf t}, \hbar) =\sum_{j_L, j_R} \sum_{w, {\bf d} } 
N^{{\bf d}}_{j_L, j_R}  \frac{\sin\frac{\hbar w}{2}(2j_L+1)\sin\frac{\hbar w}{2}(2j_R+1)}{2 w^2 \sin^3\frac{\hbar w}{2}} \re^{-w {\bf d}\cdot{\bf  t}},
 \ee
 as well as the twisted free energy
  \be \label{NS-inst-hat}
{\widehat{F}}^{\rm NS,inst}({\bf t}, \hbar) =\sum_{j_L, j_R} \sum_{w, {\bf d} } 
N^{{\bf d}}_{j_L, j_R}  \frac{\sin\frac{\hbar w}{2}(2j_L+1)\sin\frac{\hbar w}{2}(2j_R+1)}{2 w^2 \sin^3\frac{\hbar w}{2}} \re^{-w {\bf d}\cdot\left({\bf  t+\ri \pi {\bf B}}\right)},
 \ee
 \be
 \label{NS-j-hat}
{\widehat{F}}^{\rm NS}({\bf t}, \hbar) ={1\over 6 \hbar} a_{ijk} t_i t_j t_k +b^{\rm NS}_i t_i \hbar +{\widehat{F}}^{\rm NS,inst}({\bf t}, \hbar),
\ee
and the twisted quantum mirror map
 \be\label{qmirhat} -{\widehat{t}}_i(\hbar)=\log z_i+ \Pi_i( \bf z_{B}, \hbar), \ee
where ${\bf z_{B}}=(z_1 (-1)^{B_1}, \cdots, z_{g_\Sigma+r_\Sigma} (-1)^{B_{g_\Sigma+r_\Sigma}})$.

\bibliographystyle{JHEP}

\linespread{0.6}
\bibliography{biblio}
\end{document}